\documentclass[usenatbib,usegraphicx]{mn2e}
\onecolumn
\usepackage{subfigure}
\usepackage{longtable}
\usepackage{graphicx}
\usepackage{graphics}
\usepackage{keyval}
\usepackage{trig}

\def\beq{\begin{equation}}
\def\eeq{\end{equation}}
\def\bey{\begin{eqnarray}}
\def\eey{\end{eqnarray}}

\def\kpc{\, {\rm kpc} }
\def\msun{M_\odot}

\def\Msun{M_\odot}

\def\kms{\, {\rm km \, s}^{-1} }

\def\grad{{\bf \nabla}}

\def\a0{$a_0$}

\title[Milky Way potentials in CDM and MOND]{Milky Way potentials in CDM and MOND. Is the Large Magellanic Cloud on a bound orbit?}
\author[Wu et al.]{Xufen Wu$^{1}$, Benoit Famaey$^{2}$, Gianfranco Gentile$^{3}$, Hagai Perets$^{4}$ and HongSheng Zhao$^{1}$\\
$^{1}$SUPA, School of Physics and Astronomy, University of St. Andrews, Fife KY16 9SS, UK\\
$^{2}$Institut d'Astronomie et d'Astrophysique, Universit\'e Libre de Bruxelles,
CP 226, Boulevard du Triomphe, B-1050, Bruxelles, Belgium\\
$^{3}$University of New Mexico, Department of Physics and Astronomy, 800 Yale Boulevard NE, Albuquerque, NM 87131\\
$^{4}$Benoziyo centre for Astrophysics, Weizmann Institute of Science, POB 26, Rehovot, Israel }
\begin{document}



\maketitle

\label{firstpage}

\begin{abstract}
We compute the Milky Way potential in different cold dark matter (CDM) based models, and compare these with the modified Newtonian dynamics (MOND) framework. We calculate the axis ratio of the potential in various models, and find that isopotentials are less spherical in MOND than in CDM potentials. As an application of these models, we predict the escape velocity as a function of the position in the Galaxy. This could be useful in comparing with future data from planned or already-underway kinematic surveys (RAVE, SDSS, SEGUE, SIM, GAIA or the hypervelocity stars survey). In addition, the predicted escape velocity is compared with the recently measured high proper motion velocity of the Large Magellanic Cloud (LMC). To bind the LMC to the Galaxy in a MOND model, while still being compatible with the RAVE-measured local escape speed at the Sun's position, we show that an external field modulus of less than $0.03 a_0$ is needed.
\end{abstract}


\section{Introduction}

The classical problem of modelling the mass distribution and the corresponding gravitational potential of our Galaxy carries with it the key to our understanding of the elusive dark matter. Nowadays, the dominant paradigm is that dark matter is actually made of non-baryonic weakly interacting massive particles, the so-called cold dark matter (CDM). Many models of the Milky Way have been devised in this context (e.g., Wilkinson \& Evans 1999, Olling \& Merrifield 2001, Klypin, Zhao \& Somerville 2002).

However, a modification of the Newtonian dynamics (MOND) has been suggested as an alternative to those CDM models. In MOND the Newtonian gravitational acceleration $g_N$ is replaced with $g\sim\sqrt{g_{N}a_0}$ when the gravitational acceleration is far smaller than the acceleration constant  $a_0=1.2\times10^{-10}{\rm ms}^{-2}$, (e.g., Milgrom 1983a,b,c; Bekenstein \& Milgrom 1984; Sanders \& McGaugh 2002; Bekenstein 2006; Milgrom 2008; Zhao 2008). MOND tightly fits the observations without dark matter in different types of galaxies (e.g., Milgrom \& Sanders 2003; Sanders \& Noordermeer 2007; Nipoti et al. 2007; Famaey et al. 2007a; Gentile et al 2007a,b; Tiret et al. 2007; Sanchez-Salcedo et al. 2008). Moreover, the research on covariant MOND theories (e.g., Bekenstein 2004; Sanders 2005; Bruneton \& Esposito-Far\`ese 2007; Zhao 2007) now enables the study of the Cosmic Microwave Background (e.g., Skordis et al. 2006), the growth of structure (e.g., Halle et al. 2007; Skordis 2008), and gravitational lensing of galaxies (e.g., Zhao et al 2006; Chen \& Zhao 2006) and clusters of galaxies (e.g., Angus et al. 2007; Famaey et al. 2007b; Milgrom \& Sanders 2007).

The CDM and MOND interpretation of galaxy kinematics can sometimes be viewed as degenerate, at the price of accepting some mysterious conspiracy between the distribution of baryons and dark matter at all radii in the CDM context (see e.g. McGaugh 2005). There are, however, some fundamental differences: one of them is the role of dynamical friction (e.g., Sanchez-Salcedo et al. 2006; Nipoti et al. 2008). Another one is the fact that the Strong Equivalence Principle is violated in MOND when considering the external field in which a system is embedded (e.g., Bekenstein \& Milgrom 1984; Zhao \& Tian 2006; Famaey, Bruneton \& Zhao 2007c; Sanchez-Salcedo \& Hernandez 2007). Hence, it is also important to study the effects of the environment in MOND: unlike in CDM, the rotation curves and the morphology of galaxies depend on both the background and self-gravity acceleration (Wu et al. 2007). The escape velocity at any location in a galaxy is also dependent on this external field (Famaey et al. 2007c; Wu et al. 2007). It is however virtually impossible to precisely calculate the value of the external field at the location of the Milky Way without describing the Local Group formation in the context of a MOND N-body simulation. Nevertheless, one can estimate its order of magnitude from the acceleration endured during a Hubble time in order to attain a peculiar velocity of 600~${\rm km} \, {\rm s}^{-1}$ with respect to the CMB, $g_{\rm ext} \simeq H_0 \times 600 \, {\rm km} \, {\rm s}^{-1} \simeq 0.01 a_0$. This value is also {\it roughly} the one produced by the M31 galaxy, as well as the one produced by the Virgo and Coma clusters (Famaey et al. 2007c).

In this paper, we model the Milky Way with different CDM-based models, and compare them with the MOND framework, allowing for various values of the external field. We show that while the CDM models have a spherical potential in the outer regions of the Galaxy ($r=50$~kpc), the MOND potential is more ellipsoidal. 
We then also numerically calculate the circular and escape velocities from the Galaxy as a function of position in both MOND and CDM models, and finally apply this to the case of the large Magellanic cloud (LMC) in order to see which models allow it to be bound to the Milky Way. We conclude that, in MOND, an external field of less than $0.03a_0$ is needed, an upper limit strikingly similar to that needed to explain the local escape speed from the solar neighbourhood (Famaey et al. 2007c; Wu et al. 2007).

\section{A covariant framework for MOND}

There now exists several covariant frameworks for MOND (see, e.g., Bekenstein 2004; Sanders 2005; Bruneton \& Esposito-Far\`ese 2007; Skordis 2008, Halle et al. 2008, and references therein). 
In the quasi-static and weak field limit of these theories, the slightly bent metric for a galaxy is given by 
\begin{equation}\label{metric} 
g_{\alpha\beta} dx^{\alpha}dx^{\beta} 
=-(1+{2\Phi \over c^2}) d(ct)^2 + (1-{2\Psi \over c^2}) dl^2 
\end{equation} 
where $dl^2=\left(dx^2 +dy^2 + dz^2 \right)$ 
is the Euclidean distance in cartesian coordinates, and 
the Newtonian gauge is adopted with 
the potential being $\Phi(x,y,z)$ and $\Psi(x,y,z)$. 

As one of the simplest frameworks, one can adopt 
the ${\bf \nu\Lambda}$ co-variant formulation of Zhao (2007), and Halle et al.~(2008). The ${\bf \nu\Lambda}$ model works as GR, except that 
the space-time is filled with a {\bf N}on-{\bf u}niform Dark Energy fluid 
described by a four-vector ${\bf \nu}^\alpha$, with unit norm and parallel to the local time direction. If we define a dimensionless co-variant vector 
\bey 
E^\beta &\equiv& \frac{c^2 \nabla_\parallel {\nu}^\beta}{a_0} 
\sim (0, -{\partial_x \Phi \over a_0} , -{\partial_y \Phi \over a_0}, 
-{\partial_z \Phi \over a_0} ) ~\mbox{\rm near galaxies}, 
\eey 
where $\nabla_\parallel \nu = \nu^\alpha \nabla_\alpha \nu$ is
the four-vector's covariant derivative 
projected along the local time direction, we can consider the model with a vector field Lagrangian 
density given by ${\cal L}= {\cal F}(|E|^2) {a_0^2 /16 \pi G} $, where ${\cal F}$ is a dimensionless funtion of 
$|E|^2 \equiv g_{\alpha\beta} E^\alpha E^\beta$.
We then minimize against the total action which 
contains this ${\bf \nu\Lambda}$ fluid action plus the 
the usual Einstein-Hilbert action and the usual matter action. 
The equation of motion of the vector field then sets 
the vector field to be ${\bf \nu}^\alpha \sim (1-\Phi c^{-2},0,0,0)$ 
in the weak-field quasi-static limit, while the ij-cross-term of the Einstein equation requires the Newtonian potentials $\Phi$ and $\Psi$ to be equal in the absence of anisotropic stress. The tt-term and the ii-terms of the Einstein equations can thus be combined to yield:
\begin{equation}\label{tt}
-\grad \cdot [(1-{\cal F}') (-\grad \Phi)] - Q/2 = 4 \pi G \left(\rho + \frac{P}{c^2} \right) ,
\end{equation}
where $Q$ is a function of $\dot{\Phi}$ and of scalar mode perturbations of the four-vector ${\bf \nu}^\alpha$, and is negligible near collapsed quasi-static systems, like the pressure term is.   
The function $1-{\cal F}'$ is a dielectric-like scalar free function of the covariant scalar quantity $|E|=|\grad \Phi|/a_0$. This dielectric-like function is thus the equivalent of the $\mu$-function of MOND (Bekenstein \& Milgrom 1984). To recover the MOND phenomenology, a specific choice of the function can be
\beq
{\cal F}' \equiv \frac{{\rm d}{\cal F}}{{\rm d}{|E|^2}}= \left(1 + {|E| \over n}\right)^{-n}.
\eeq
Zhao (2007) prefers the parameter $n \sim 3$ to best explain the amplitude of the cosmological constant.  Here we rather choose $n=1$ corresponding to the `simple' $\mu$-function of MOND (Famaey \& Binney 2005; Zhao \& Famaey 2006; Famaey et al. 2007a; Sanders \& Noordermeer 2007). Note finally that, when a galaxy is embedded in an external field $ {\bf g}_{\rm ext}$, we have that the {\it total} gravitational force $-\nabla \Phi \equiv {\bf g}_{\rm ext} - {\mathbf\grad} \Phi_{\rm int}$, where $\Phi_{\rm int}$ is the internal potential of the galaxy.

\section{Potential of the Milky Way}

\subsection{MOND}

We model the Milky Way in MOND following Wu et al.~(2007), where the Besan\c{c}on baryons model (Robin et al. 2003) is used. This model contains a thin stellar disc with mass of $2.15\times 10^{10}\Msun$, a thick stellar disc of $3.91\times 10^9 \Msun$, a bar/bulge of $2.03\times 10^{10}\Msun$, an interstellar gaseous matter component of $4.95\times 10^9\Msun$, and a stellar halo of $2.64\times 10^8 \Msun$. We first choose an external field $g_{\rm ext} = 0.01 a_0$: this is presumably due to the 
local gravitational attraction of Large Scale Structure, and mainly from the so-called Great Attractor region (see Radburn-Smith et al. 2006) in the Sun-Galactic Centre direction. We then let the external field vary, in order to investigate the effect of its value on the MOND model. Indeed, the effect of M31, and of the Coma and Virgo clusters, makes the value of the external field extremely uncertain. For instance, the baryonic mass of M31 is estimated to lie between $7\times10^{10}M_\odot$ and $2\times10^{11}M_\odot$ (Seigar et al. 2006; Tempel et al. 2007; Geehan et al. 2006; Carignan et al. 2006), and sits at 800~kpc from the Milky Way center, thus exerting an external field in the range $0.01a_0-0.02a_0$, a value roughly similar to the one exerted by Large Scale Structure. Note however that the external field from M31 is a varying one, and that if one {\it only} considered the external field from M31, no star could ever escape from the MW-M31 system because the external field from M31 would {\it never} dominate over the internal one from the MW in the direction opposite to M31. A star from the solar neighbourhood leaving in this direction would thus never escape: this is why the external field from Large Scale Structure is also important. In any case, we consider hereafter the external field as a parameter that can be constrained by data (such as the proper motion of the LMC, see Sect.~5.2).

Once the external field is chosen, we solve for the internal gravitational acceleration, $- {\mathbf\grad} \Phi_{\rm int}$, the modified Poisson equation of MOND (see also Eq.~3) 
\beq
-\nabla \cdot \left[ \mu(X) ({\bf g}_{\rm ext} - {\mathbf\grad} \Phi_{\rm int}) \right]=4\pi G\rho,\qquad X={|{\bf g}_{\rm ext} - {\mathbf\grad} \Phi_{\rm int}| \over a_0} 
\eeq with the boundary condition 
\beq - {\mathbf\grad} \Phi_{\rm int} \rightarrow - {\mathbf\grad} \Phi_{\rm int}^{\infty}(x',y',z') 
\eeq on all the most external grid points (labelled $(x',y',z')$ here), and 
\beq \Phi_{\rm int}^{\infty}=-{GM\over \mu(g_{\rm ext}/a_0) \sqrt{(1+\triangle)(y^2+z^2)+x^2+s^2}}, 
\eeq where $\triangle=\left({d\ln\mu/ d\ln X}\right)_{X=g_{\rm ext}/a_0}$ is a dilation factor in the range [0,1] (Bekenstein \& Milgrom 1984; Milgrom 1986a; Zhao \& Tian 2006; Zhao \& Famaey 2006), and $s$ is a softening factor comparable to the half-light radius of the galaxy. 

As stated hereabove, we choose the `simple' $\mu$-function of the form $\mu(X)= X/(1+X)$  (Famaey \& Binney 2005; Zhao \& Famaey 2006; Famaey et al. 2007a; Sanders \& Noordermeer 2007), corresponding to $n=1$ in Eq.~(4). We use the Poisson solver developed by the Bologna group (Ciotti et al. 2006) with $512\times 64\times 128$ grid points, the radial grid points being $r_i=50.0 \tan \left[(i+0.5){0.5\pi /512+1}\right]$~kpc. 
Notice that, even though it has less degrees of freedom than the CDM models presented hereafter, the MOND model has, contrary to common wisdom, some freedom: the choice of the $\mu$-function could have been different, the value of $a_0$ has an error bar that we ignored here, the mass-to-light ratio of the stellar component can vary, the presence of unseen baryons can have a big influence (see Sect.~5.4), and the value of the external field is not known a priori.

We note that, in such a MOND model, the escape velocity $v_{\rm esc}$ of a test particle at an arbitrary position $(x,y,z)$ can be defined as 
\beq {v_{\rm esc}^2(x,y,z)\over 2}+\Phi_{\rm int}(x,y,z)\equiv E_{\rm eff}=0.\eeq This means that for a zero effective internal energy of the particle ($E_{\rm eff} = 0$), it can escape from the internal system and never return.

\subsection{CDM}

We now compare the MOND model described above with various CDM models of the Galaxy. Unfortunately, CDM-based models are far from unique. Nevertheless, we limit ourselves to four models: two based on the work of Klypin, Zhao \& Somerville (2002, hereafter KZS), and two based on the recent work of the RAVE collaboration on the local escape speed from the Solar Neighbourhood (Smith et al. 2007). 

First we explore the model labelled B1 in KZS, composed of a double-exponential disc for the baryons and a NFW profile (Navarro et al. 1997) for the dark matter halo. We then explore another model in which we replace the KZS baryons with the Besan\c{c}on model, but use the same CDM component as in KZS B1. We call this second model the Bsc CDM model.

The NFW profile is the most widely used CDM-halo model, it is described as (Navarro et al. 1997)
\bey &\rho_{halo}(r) &={\rho_s \over x(1+x)^2}, x=r/r_s,\\
&M_{vir}&={4\pi \over 3}\rho_{cr}\Omega_0\delta_{th}r_s^3c^3,\\
&\rho_{s}&={\rho_{cr}\Omega_0\delta_{th}\over 3}{c^3\over \ln (1+c)-{c\over(1+c)}}.
\eey
Where $\rho_s$ is the characteristic density parameter, $r_s$ is the radius parameter, $\rho_{cr}=3H_0^2 / 8\pi G$ is the critical density of the Universe determined by the Hubble constant at redshift $z=0$, $\Omega_0$ is the fraction of matter (including baryons and dark matter) to the critical density, $\delta_{th}$ is the critical overdensity of the virialized system, and $c$ is the concentration parameter. In the inner part of the dark halo, $\rho_{halo}\propto r^{-1}$ and in the outer part $\rho_{halo}\propto r^{-3}$.

\begin{table*}
\caption{NFW Profile Parameters}
\begin{tabular}{lcccccccc}
\hline
$NFW$ & $c$ & $M_{vir}(10^{12}\Msun)$ & $r_{vir}$(kpc) & $\delta_{th}$ & $H_0$($\kms$ Mpc$^{-1}$) & $\Omega_0$ & $r_{\odot}$(kpc) \\
\hline
$KZS B1$ & 12 & 1.0 & 258 & 340 & 70 & 0.3 & 8.5\\
$RAVE1$ & 24.3 & 1.89 & 257 & 340 & 65 & 0.3 & 7.5\\
\hline
\end{tabular}
\end{table*}

The NFW profile parameters of the KZS B1 model are listed in Table 1. Both angular momentum exchange between the baryons and dark matter and adiabatic contraction are considered. The angular momentum of baryons is lost and deposited into the dark halo, hence the centre of the dark halo becomes more scattered. During the evolution of the galaxy, the fall of baryons into the galactic centre makes the gravity potential deeper, and more dark matter particles are trapped in this deeper potential well. The adiabatic contraction thus makes the dark halo become denser. These two conflicting effects have been shown to make the CDM halo approximately similar to the original NFW-profile. For baryons, the KZS B1 model has a double-exponential disc. The density of nucleus, bulge and disc is then described by  \footnote{Eq.~6 of KZS had a typo} (Kent et al. 1991; Zhao 1996):
\bey &\rho_b&=\rho_1+\rho_2+\rho_3 ,\\
&\rho_1&=\rho_{1,0}\left({0.36(x^2+y^2)+z^2 \over z_0^2}\right)^{-1.85/2} \exp\left(-\sqrt{0.36(x^2+y^2)+z^2 \over z_0^2}\right) ,\\
&\rho_2&=\rho_{2,0}\exp\left(-{\sqrt{[(0.26x)^2+(0.69y)^2]^2+z^4} \over 2z_0^2}\right) ,\\
&\rho_3&=\rho_{3,0}\exp\left(-{\sqrt{x^2+y^2}+12 |z| \over r_d}\right)
\eey
where $z_0=0.4$~kpc, $r_d=3$~kpc, $M_1+M_2=1\times10^{10}\Msun$, $M_1/(M_1+M_2)=0.15$, $M_3=5\times 10^{10} \Msun$, and $\rho_{1,0},\rho_{2,0},\rho_{3,0}$ are characteristic parameters for the density required in order to reproduce the bar/bulge from COBE, and the disc of Kent et al. (1991). The three densities $\rho_1$,$\rho_2$,$\rho_3$ are convenient for describing the global shape of the Galaxy rather than real separate components of the MW. This is why we also set up the Bsc model as explained above.

As we aim at predicting the escape velocity as a function of position, we explore additional models that were used by the RAVE collaboration to reproduce the local escape velocity from the Solar Neighbourhood (Smith et al. 2007). We choose two of their models, the uncontracted NFW (RAVE~1) and the Wilkinson-Evans (WE) profile (RAVE~3) for the dark halo.

Model RAVE~1 is described by the combination of a NFW halo (Parameters in Table 1), a baryonic disc with a Hernquist bulge of $1.5\times 10^{10}\Msun$ and a scale radius of $0.6$~kpc, and a Miyamoto-Nagai disc of $5.0\times 10^{10}\Msun$ with scale height $0.3$~kpc and scale length of $4$~kpc.

In the RAVE~3 model the dark matter halo density profile is more cuspy in the centre, $\rho \sim r^{-2}$, and has a cut-off in the density distribution at the outer parts of the Galaxy, $\rho \sim r^{-5}$. The halo density is given by (Wilkinson \& Evans~1999)
\beq 
\rho(r)={M\over 4\pi}{a_s^2\over r^2(r^2+a_s^2)^{3/2}},
\eeq
where $M=1.89\times 10^{12}\Msun$ is the total mass of the halo, and $a_s=314$~\kpc.  This profile keeps the rotation curve approximately flat up to the radius $a_s$.

\section{Axis ratios and Phantom dark matter}

\begin{figure}{}
\resizebox{16cm}{!}{\includegraphics{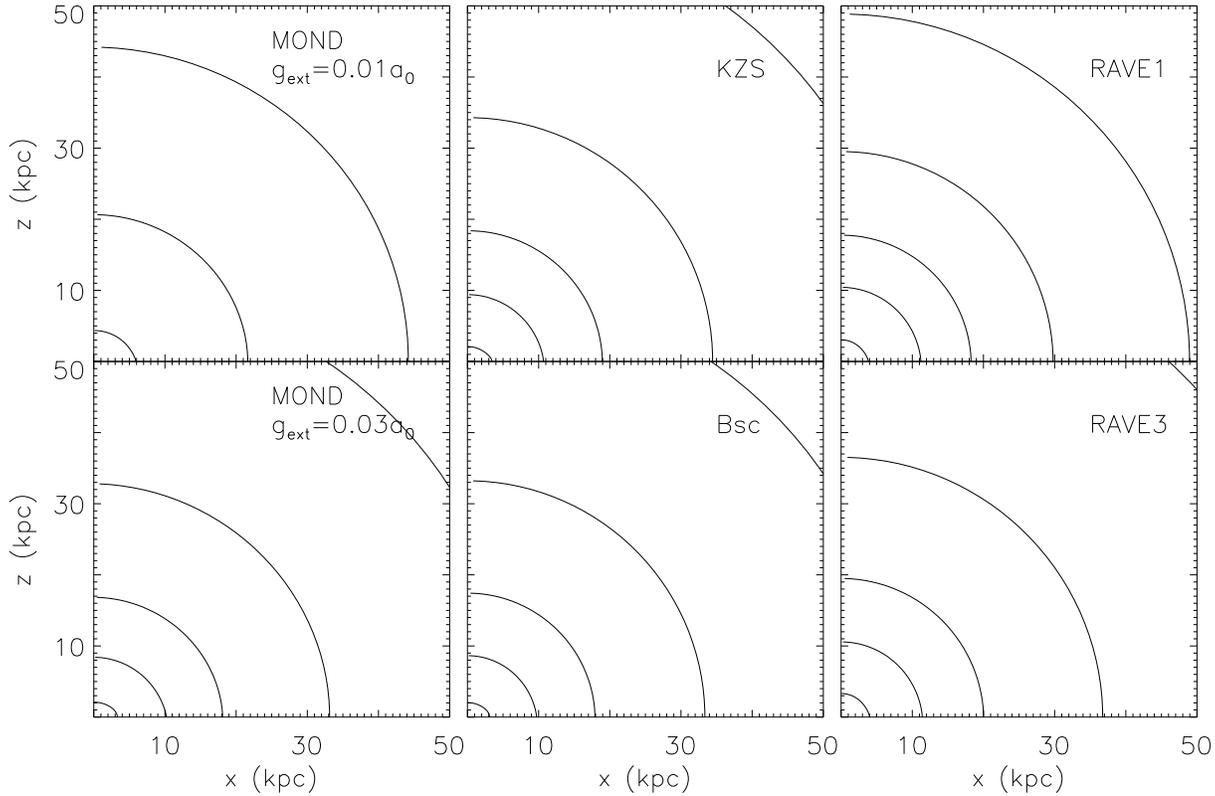}}\vskip 0.5cm
\caption{Isopotentials in the $x-z$ plane. The $x$-axis is sun-the centre direction and the $z$-axis is perpendicular to the disc plane. The isopotentials correspond to escape velocities of 600~km/s, 500~km/s, 450~km/s, 400~km/s and 350~km/s (starting from 600~km/s in the center).}\label{mondpot}
\end{figure}

\begin{figure}{}
\resizebox{8cm}{!}{\includegraphics{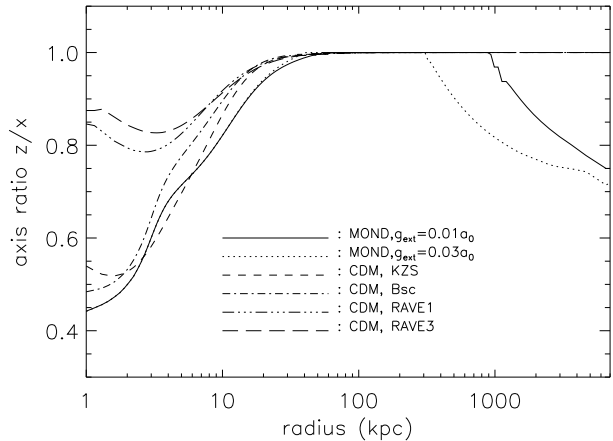}}\vskip 0.5cm
\caption{$z-x$ potential axis ratio. The $x$-direction is the galactic centre to sun direction, and $z$  is perpendicular to the disc plane. The predictions for $r>500$~kpc are only qualitative because they should be perturbed by variations of the external field and covariant corrections.}\label{ratio}
\end{figure}

\begin{figure}{} 
\resizebox{16cm}{!}{\includegraphics{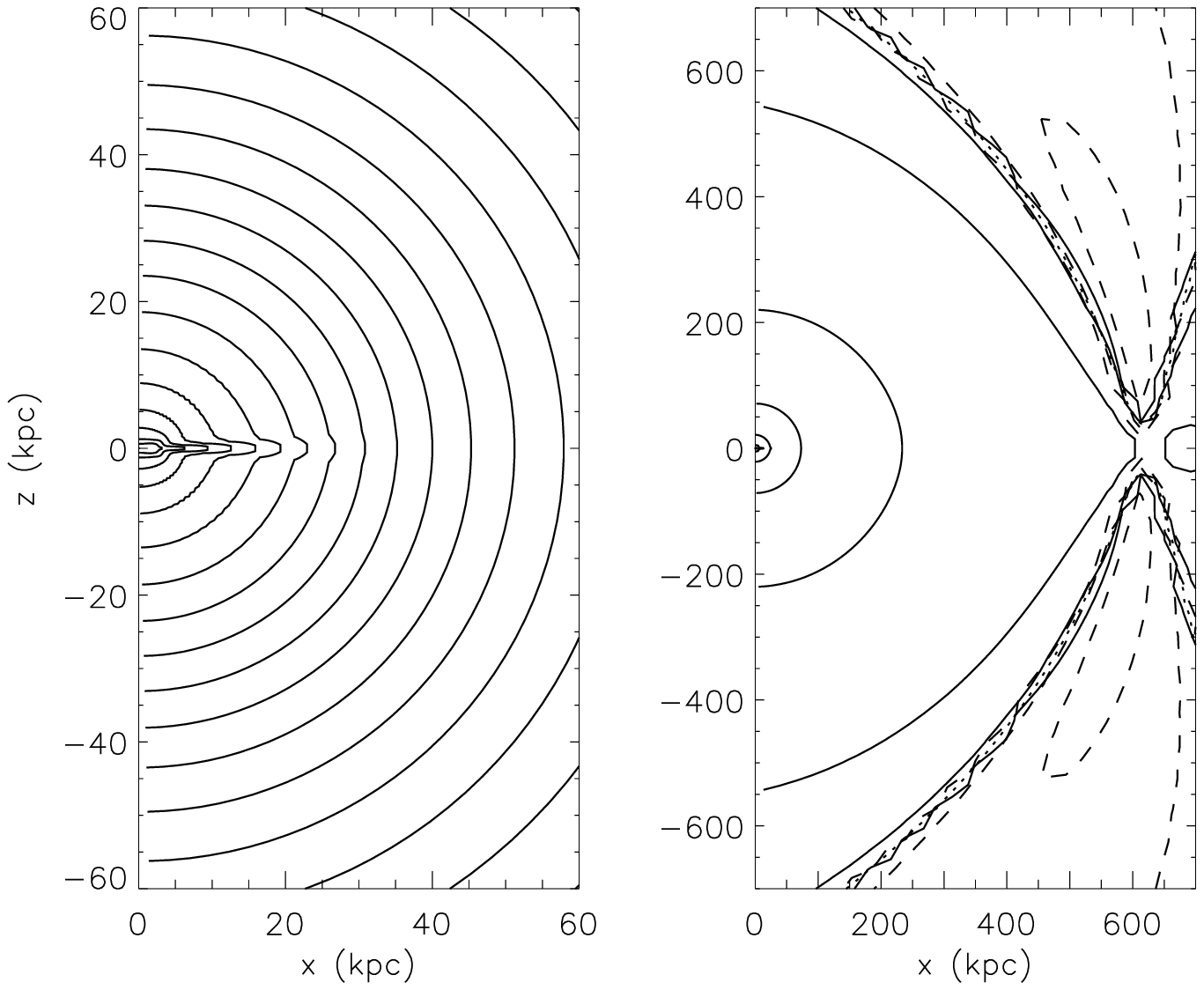},\includegraphics{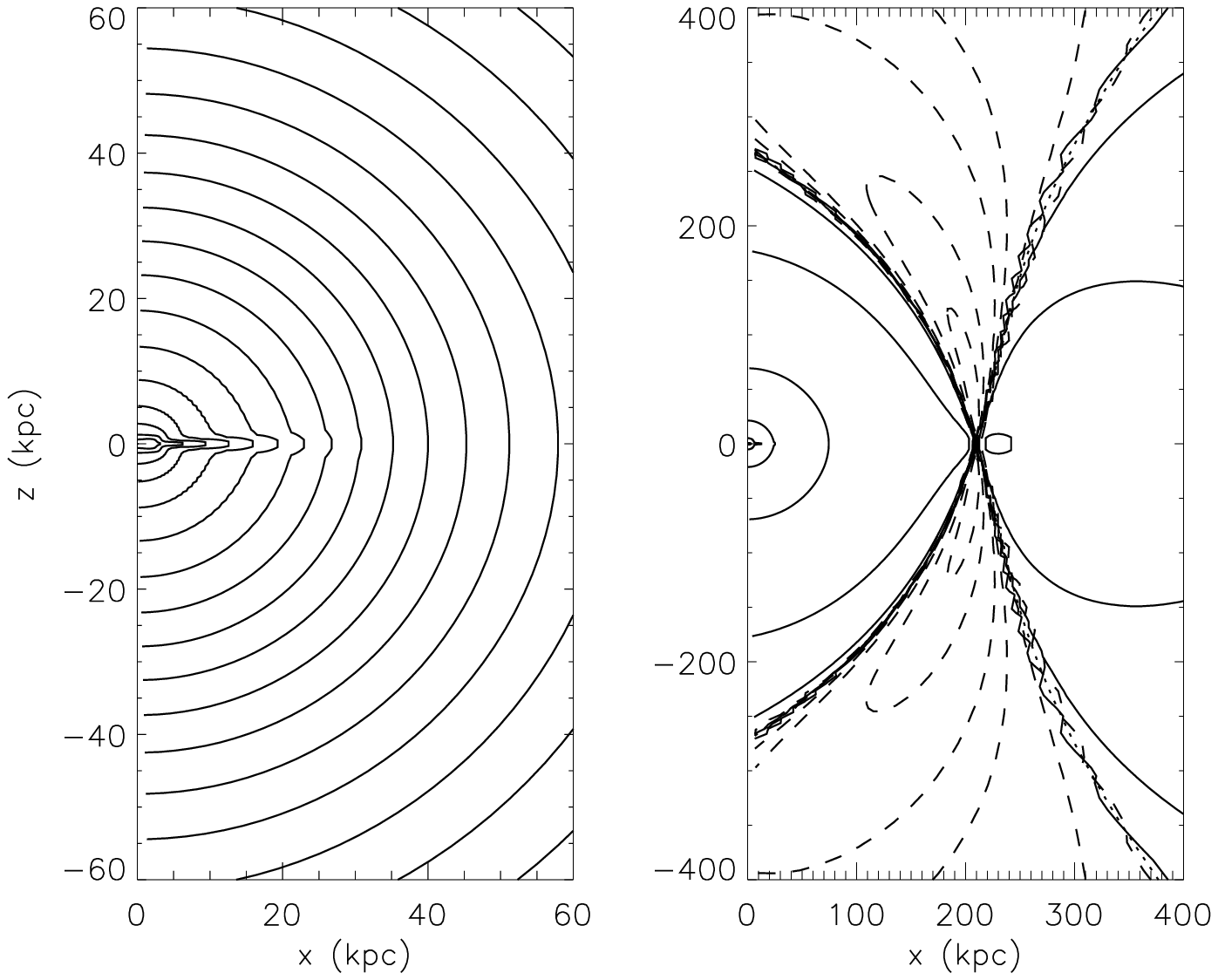}}\vskip 0.5cm 
\caption{Isodensity of phantom dark matter in a MOND Milky Way embedded in an external field of $0.01a_0$(panels a and b) and $0.03a_0$(panels c and d). The 
solid and dashed contours are isodensity for positive and negative density respectively, and the dotted line is the watershed with zero-density.}\label{effden} 
\end{figure} 

To investigate the intrinsic differences between the MOND and CDM potentials, we compute and plot the isopotential contours of the MOND model with an external field of $0.01a_0$ and $0.03a_0$, and similar contours of the four CDM models on Fig~\ref{mondpot}. The MOND potential always yields slightly more ellipsoidal (oblate) isopotential contours. This is mainly because the MOND potential is produced by baryons and the flattened/non-axisymmetric density distribution of the disc/bar in the Besan\c{c}on model dominates the shape of the potential at small radii. Moreover, we note that when an external field is applied in a random direction (here it has been applied in the $x$-direction), even an axisymmetric baryonic density distribution yields a triaxial/non-axisymmetric potential \footnote{This inherent triaxiality of the internal potential in MOND  is also mentioned in Wang et al. (2008)}. Here, the external field in the $x$-direction actually makes the MOND potential oblate again at very large radii (see Eq.~7) asymptoting to a $z-x$ axis ratio of 0.7 at infinity (see Fig.~\ref{ratio}). The MOND potentials have thus a flattened oblate shape at the very edge of the Galaxy too. Note that our predictions at large radii ($r> 500$~kpc) on Fig.~2 are only qualitative since the MOND potential at extremely large radii can of course become dominated by other local structures in certain directions, and there can also be variations of the Large Scale Structure-external field on scales larger than $1000\kpc$. Here, we only considered a constant external field. 

In the CDM models the situation is slightly different: the potential becomes spherical quite quickly as the radius increases (since it is dominated by the spherical profile of dark matter). All the CDM models have a larger $z-x$ axis ratio than the MOND ones when $r>5$~kpc. This is due only to the flattened baryonic distribution in MOND, and not to the external field effect, which becomes effective in this context only at very large radii ($r\sim 200$~kpc in the case of a $0.03a_0$ external field, see Fig.~2). Of course, assuming some flattening of the CDM halo could also yield a more oblate or or more prolate CDM potential (Jing \& Suto 2002; Bailin \& Steinmetz 2005; Allgood et al. 2006; Andrea et al. 2007), again adding to the high degree of freedom in CDM models. The MOND prediction is much more constrained. 

To investigate further the intrinsic differences between the MOND and dark matter models, it is useful to think of MOND in dark matter terms. Once the MOND potential is known, one can use the Newtonian Poisson equation to derive the corresponding density of matter that would be needed in the Newtonian context. Then, subtracting the visible matter, one obtains the ``virtual" dark matter, or ``phantom dark matter" distribution predicted by MOND. This is plotted on Fig.~\ref{effden}: clearly one sees that at small radii, the phantom dark matter tracks the baryons and effectively creates a dark matter disk (see also Nipoti et al. 2007). Moreover, cones of {\it negative} phantom dark matter densities perpendicular to the external field  direction are predicted at large radii, when the internal and external gravitational fields are of the same order of magnitude (see also Milgrom 1986b). That is one of the most important differences between MOND and CDM, since CDM of course never predicts negative mass densities. Note that this is very different from a simple change of the direction of the force felt by a test particle: here, it really means that the {\it divergence} of the force field can be locally positive, or that the flux of the force field through an infinitesimal volume is locally positive. In standard gravity, the divergence of the force field is zero outside of matter and is always negative inside the matter, through Poisson equation. A locally positive divergence of the force field is thus inconceivable in standard gravity without resorting to negative dark matter. For $g_{\rm ext}=0.01a_0$, the negative phantom dark matter region is around $r=600$~kpc while for $g_{\rm ext}=0.03a_0$ it is around $r=200$~kpc. Let us however note once again that the situation here is quite idealistic since we only took into account a constant external field. The varying external field of M31 and the LMC could create other pockets of negative phantom densities not seen here. Moreover there could also be significant covariant corrections to MOND (cf. $Q$ in Eq.~3), and variations of the LSS-external field, on scales larger than $1000\kpc$, where the predictions should thus be treated as qualitative only. We cannot observe such negative 
dark matter effects through satellite orbits, since the global shape of the MOND potential keeps the gravity tightly fitting the observations. However, one could detect such negative dark matter effect through weak gravitational lensing. In covariant MOND theories such as TeVeS (Bekenstein 2004) and the ${\bf \nu\Lambda}$ formulation (Zhao 2007, Halle et al. 2008), lensing has been shown to work {\it exactly} as in General Relativity, notably for the relation between the potential and the convergence. The only difference lies in the relation between the potential and the true underlying matter density (see Bekenstein 2004; Zhao 2007; and Angus et al. 2007 for an application). This means that, if interpreting the MOND potential in terms of dark matter as an observer believing to live in a CDM Universe would do, the phantom surface density can directly be translated into convergence: the negative dark matter would thus produce a negative convergence $\kappa$. If we placed the Milky Way far away, say $z=0.3$, then we could in principle observe the negative $\kappa$; we estimate it to be of the order of 
\beq
-\kappa \sim 0.01 {1000\kpc \over R}.
\eeq 
That is not a science-fiction story since we can look for lenses of Milky Way-like galaxies far away and the gravitational lensing should differ from CDM halo predictions. Note that there are other covariant MOND-like theories that do not predict a one-to-one relation between gravitational lensing and effective surface density (Capozziello et al. 2006, Sobouti 2007, Mendoza et al. 2007). These theories are however not, strictly speaking, true covariant formulations of MOND, notably because they do not predict this one-to-one relation (see also the discussion in Sect.~3.A of Famaey et al.~2007a). Note that the fact that such negative convergence has never been observed is not {\it per se} a falsification of TeVeS or other true covariant MOND theories, since we stress again that such an effect would be very sensitive to the detailed distribution of baryonic matter in the environment of the galaxy. We only stress that, {\it in principle}, it would not be impossible to observe it within the MOND context.

We finally note that, apart from these subtle (but important) differences, the amplitude of the total potential in the MOND model with an external field of $0.03a_0$ is very similar to the ones in the CDM models  (see Fig.~\ref{mondpot}). This is especially true for the Bsc CDM model, which shares the same baryonic distribution as the MOND models. With an external field of  $0.03a_0$, the ratio of total baryonic mass to total phantom mass created by MOND is about 0.03 at large radii, which is similar to the ratio of baryonic mass to virial mass of 0.04 in the Bsc model. This similarity 
will be even more apparent when comparing the corresponding circular and escape velocity curves in these models. Note, however, that as a consequence of the inherently flattened/triaxial potential at large radii in MOND, the escape velocity in MOND (see next section) is not completely the same at different positions on a spherical shell or cylinder of a given radius. 

\section{Circular and escape velocities}

\subsection{Prediction for galactic surveys}

\begin{figure}{}
\resizebox{16cm}{!}{\includegraphics{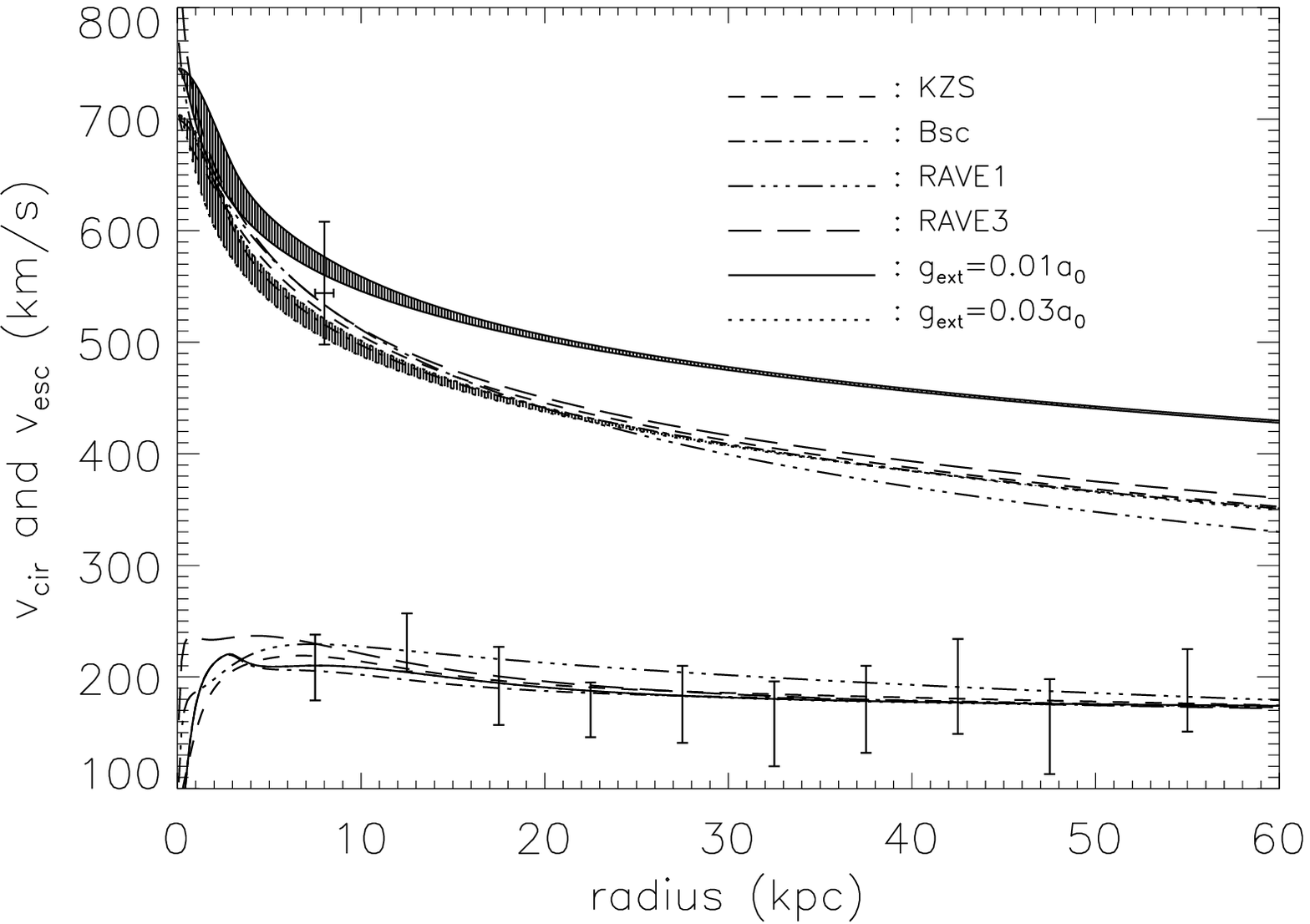},\includegraphics{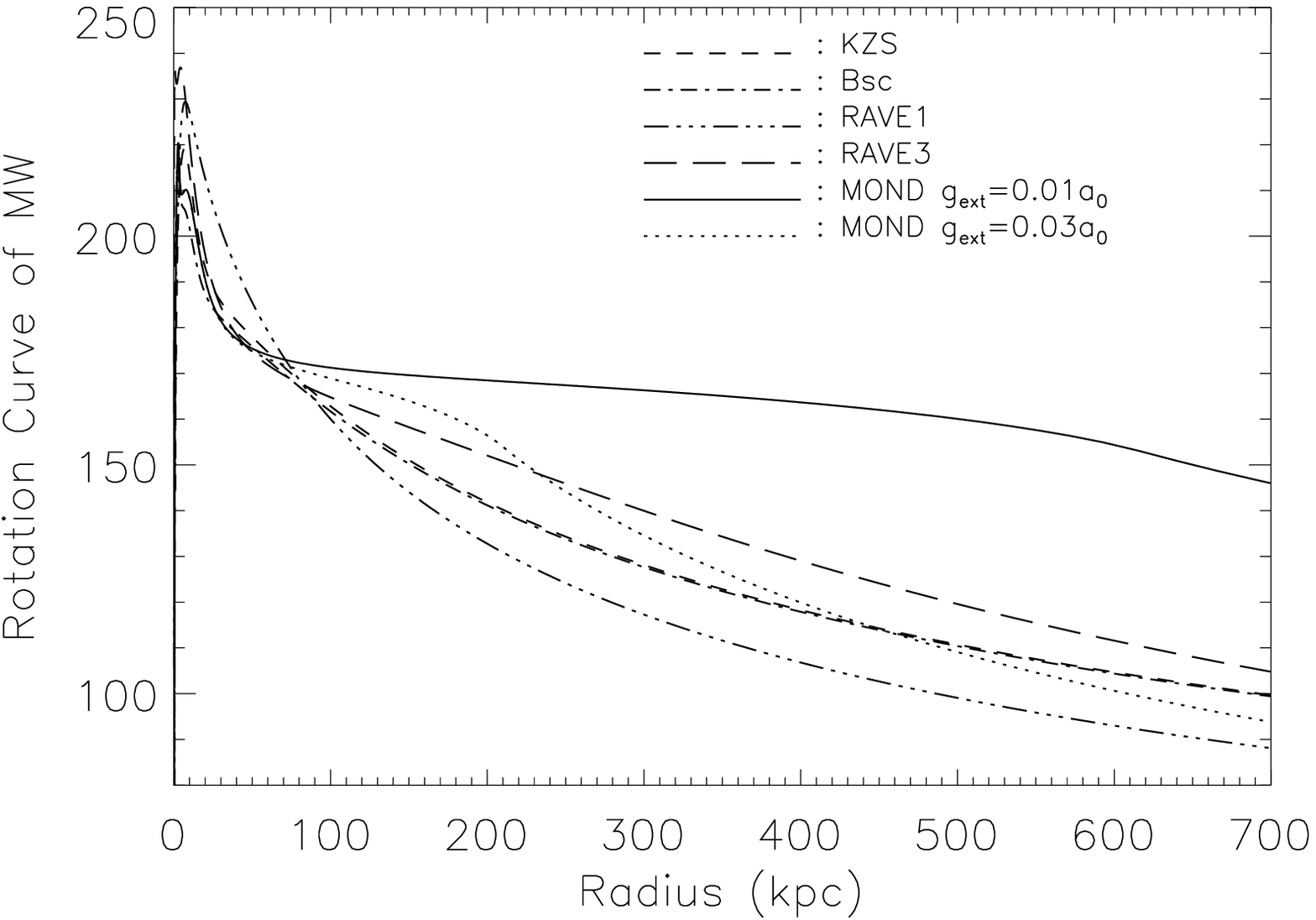}}\vskip 0.5cm
\caption{ Left panel: the circular and escape velocities in the Galaxy calculated in the different models. The two shaded areas are the escape velocity in MOND within an external field of $0.01a_0$ and $0.03a_0$. The other upper curves are the CDM predictions for the escape velocity, as shown in the label. The error bar at 7.5-8.5kpc is the RAVE escape velocity. The data below are the circular speed measured by Xue et al. (2008) from SDSS field stars data. The circular speed curves in all our models are overplotted. Right panel: The rotation curve is plotted on a very large scale, showing the return to a Keplerian behaviour at large radii, }\label{sdss}
\end{figure}

We can now compare the predicted circular and escape velocities as a function of position in the Galaxy in both the MOND and CDM context. 

The rotation curve of the Milky Way at large radii is very poorly known (see e.g. Binney \& Dehnen 1997). However, in the CDM context, Xue et al.~(2008) have recently used a set of halo stars from the Sloan Digital Sky Survey (SDSS) as kinematic tracers to estimate the rotation curve of the Galaxy out to 60~kpc. This estimated rotation curve is plotted on Fig.~4 together with the circular speed predicted in our different models. Clearly, it cannot be used to discriminate between CDM and MOND models: the escape velocity is a better discriminating test.

The MOND escape velocity as a function of radius is also plotted as shaded areas in Fig.~4 for a weak external field of $0.01 a_0$ and for an external field of $0.03 a_0$. This prediction is unique once the modulus and direction of the external field are fixed. However, the escape velocity is slightly different when considering it along different axes, which explains the use of shaded areas. Apart from the estimate based on the peculiar velocity of the Milky Way with respect to the CMB (see Sect.~1), the only real observational constraint we have on the external field modulus is the escape velocity from the Solar Neighbourhood as measured by Smith et al.~(2007) with the high velocity stars from the RAVE survey. This local escape velocity was measured to be in the range $498-608$~$\kms$ with 90\% confidence, and a median of $544$~$\kms$. This is consistent with all four CDM models presented above (by construction in the case of the two RAVE models) as can be seen on Fig.~4. To be consistent with the forementioned Besan\c{c}on MOND model, the maximum allowed modulus of the external field is $0.03 a_0$ (see also Famaey et al. 2007c; Wu et al. 2007). 

Very clearly, a higher escape velocity than the one of CDM models can actually be achieved at all radii in the most natural MOND model where the external field is $H_0 \times 600 \, {\rm km} \, {\rm s}^{-1} \simeq 0.01 a_0$. Despite their many degrees of freedom, the CDM-based models cannot achieve such high escape velocity as the MOND low external field model. Another striking prediction is that the circular and escape velocity in the MOND model with an external field of $0.03 a_0$ are extremely similar to CDM, and especially to the Bsc model. Consequently, this MOND potential effectively mimicks the NFW potential advocated by KZS as far as circular and escape velocities are concerned (but it is not strictly identical since it tracks the baryons at small radii, and is flattened by the external field at large radii, where it also predicts negative convergence for gravitational lensing).

The prediction for the escape velocity as a function of position could be highly interesting when compared with the inferred escape velocities that would be determined at various distances with planned or already-underway surveys such as RAVE, SDSS, SEGUE, SIM or GAIA.  Similarly, future observations of hypervelocity stars (Hills 1988; Brown et al. 2007) in the Galactic halo could also constrain the Galactic potential (Gnedin et al. 2005; Yu \& Madau 2007) at these large distances. High enough escape velocities inferred from such surveys could ultimately rule out the four CDM models explored here, as well as the MOND model with an external field of $0.03a_0$. 

In fact, an indication that the escape velocity might actually be higher than predicted by CDM at large radii is already suggested from the recently measured 3D velocity of the Large Magellanic Cloud (Kallivayalil et al. 2006a). As we shall see in the following, under the assumption that the LMC is bound to the Galaxy, the four CDM models presented here are only marginally viable (although not excluded).

\subsection{Escape velocity of the LMC}

\begin{table*}
\caption{Velocities and Positions of Magellanic Clouds}
\begin{tabular}{ccccc}
\hline
$Parameters$ & $Radius(x,y,z) (kpc)$&$|r|(kpc)$ & $3D-Velocity(v_x,v_y,v_z) (\kms)$ & $|v| (\kms)$\\
\hline
$LMC$ & (-0.8,-41.5,-26.9)& 49.5 & $(-86\pm12, -268\pm11, 252\pm16)$ &$378\pm18$ \\
$SMC$ & (15.3, -36.9, -43.3) & 58.9 & $(-87\pm48, -247\pm42, 149\pm37)$ & $302\pm52$\\
\hline
\end{tabular}
\end{table*}

\begin{figure}{}
\resizebox{16cm}{!}{\includegraphics{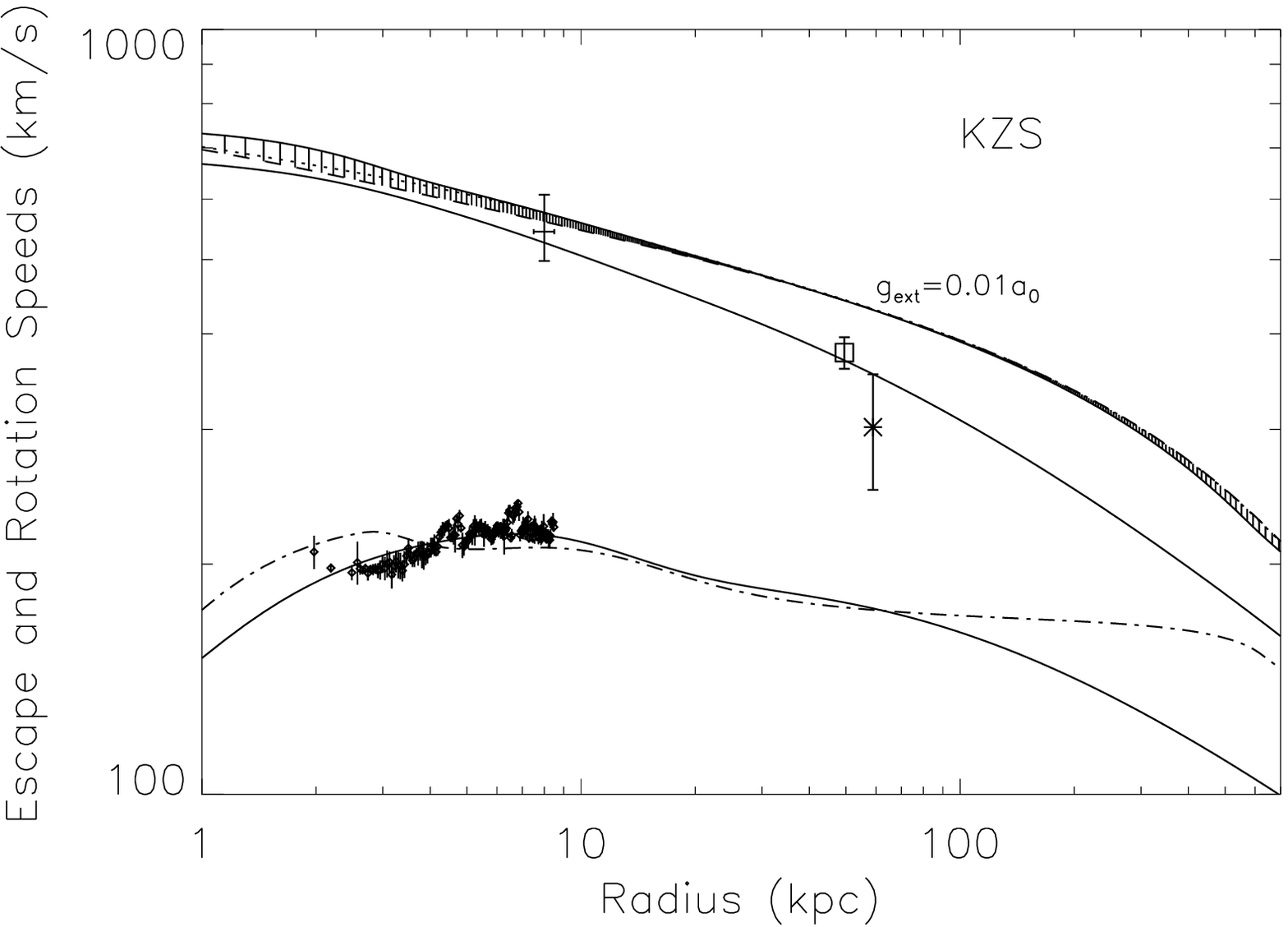},\includegraphics{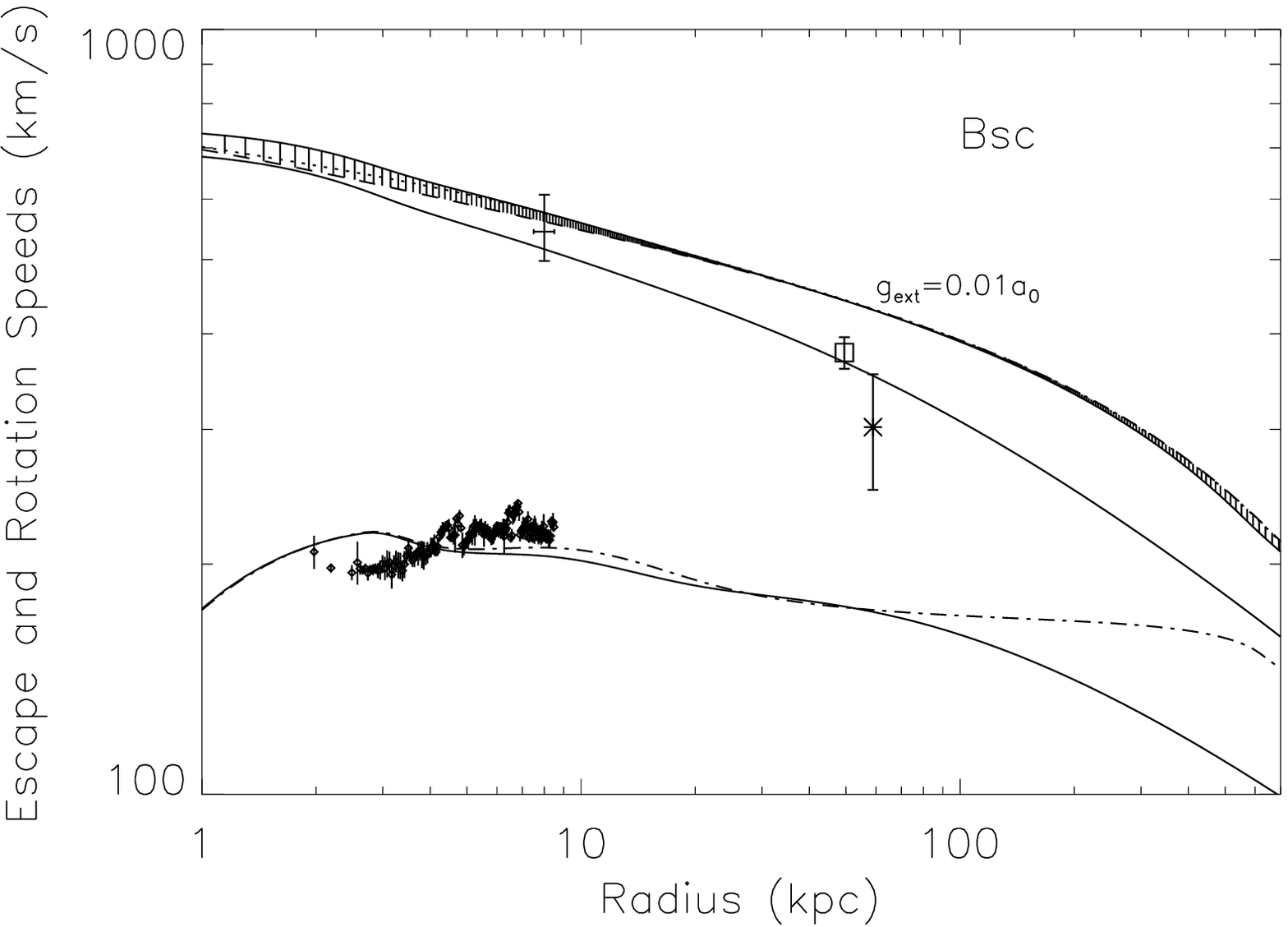}} \vskip 0.5cm
\caption{The solid lines on the left panel are the escape and circular speeds for the KZS B1 model(top solid line: escape velocity 
for MOND in the $x$-direction with external field of $0.01a_0$; Middle 
solid line: escape velocity of the KZS B1 model; Bottom solid line: circular speed of the KZS B1 model), and the solid lines on the right panel are the same for the Besan\c{c}on model combined with the same NFW profile. The dot-dashed lines are the Besan\c{c}on MOND (with low external field of $0.01a_0$) circular speed, and the shaded area is the escape velocity in MOND (with low external field of $0.01a_0$). The solid, dotted, dashed lines in the shaded area are the escape velocities in respectively the $x$ (sun to galactic centre), $y$ and $z$ (perpendicular to the disc plane) directions. Diamonds with error bars are the observed circular velocity (Clemens 1985) and the error bar at $8.0\pm0.5$~kpc is the escape velocity in the solar neighbourhood measured from the RAVE survey. The square with error bar is the newly mesaured LMC speed (Kallivayalil et al. 2006a), while the star with error bar is the SMC speed (Kallivayalil et al. 2006b).  Note that, as seen on Fig.~4, all rotation curves are also in accordance with the rotation curve of Xue et al.~(2008), not plotted in order not to overcrowd the figure.}\label{kzsbsc}
\end{figure}

\begin{figure}{}
\resizebox{16cm}{!}{\includegraphics{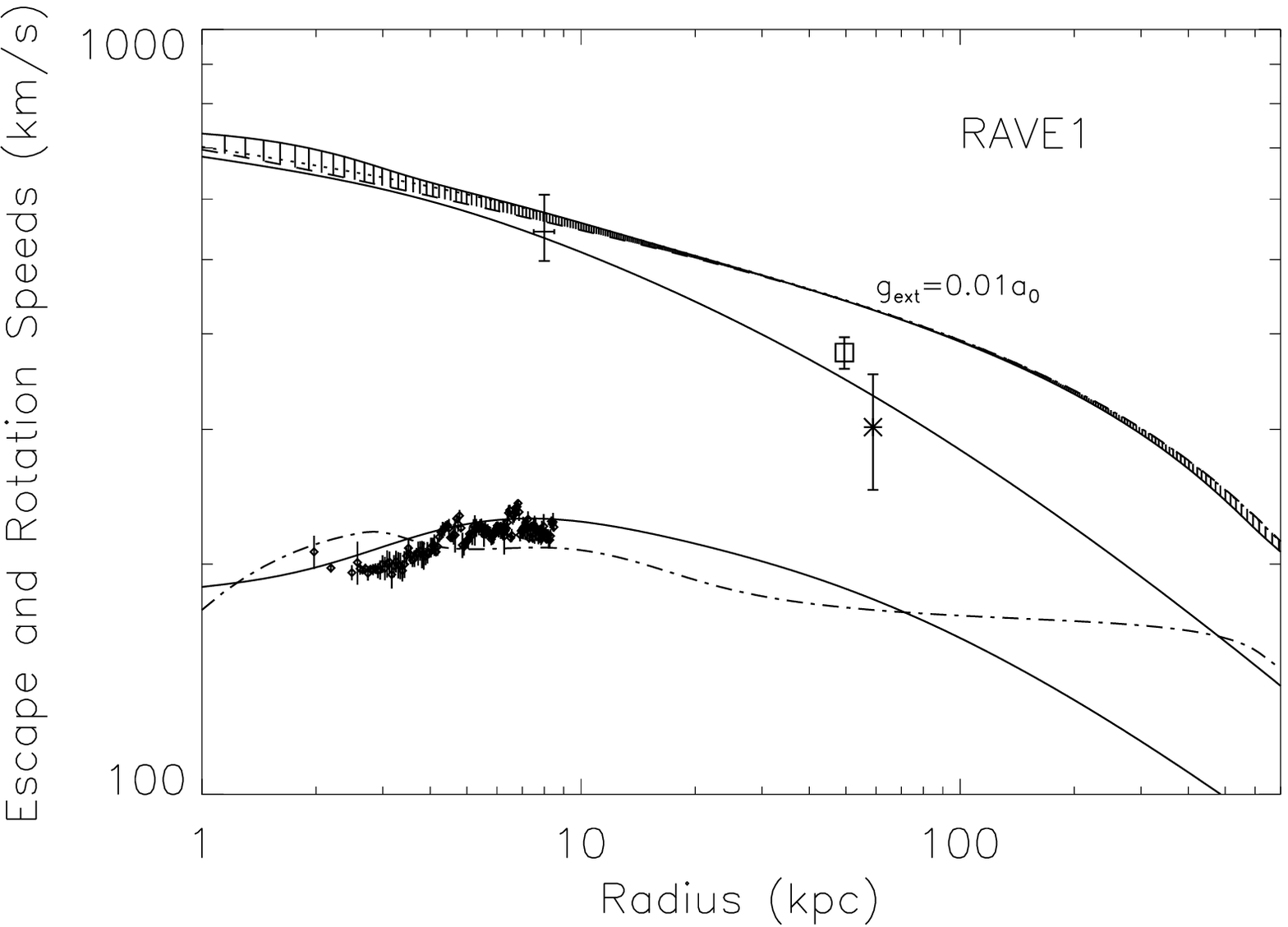},\includegraphics{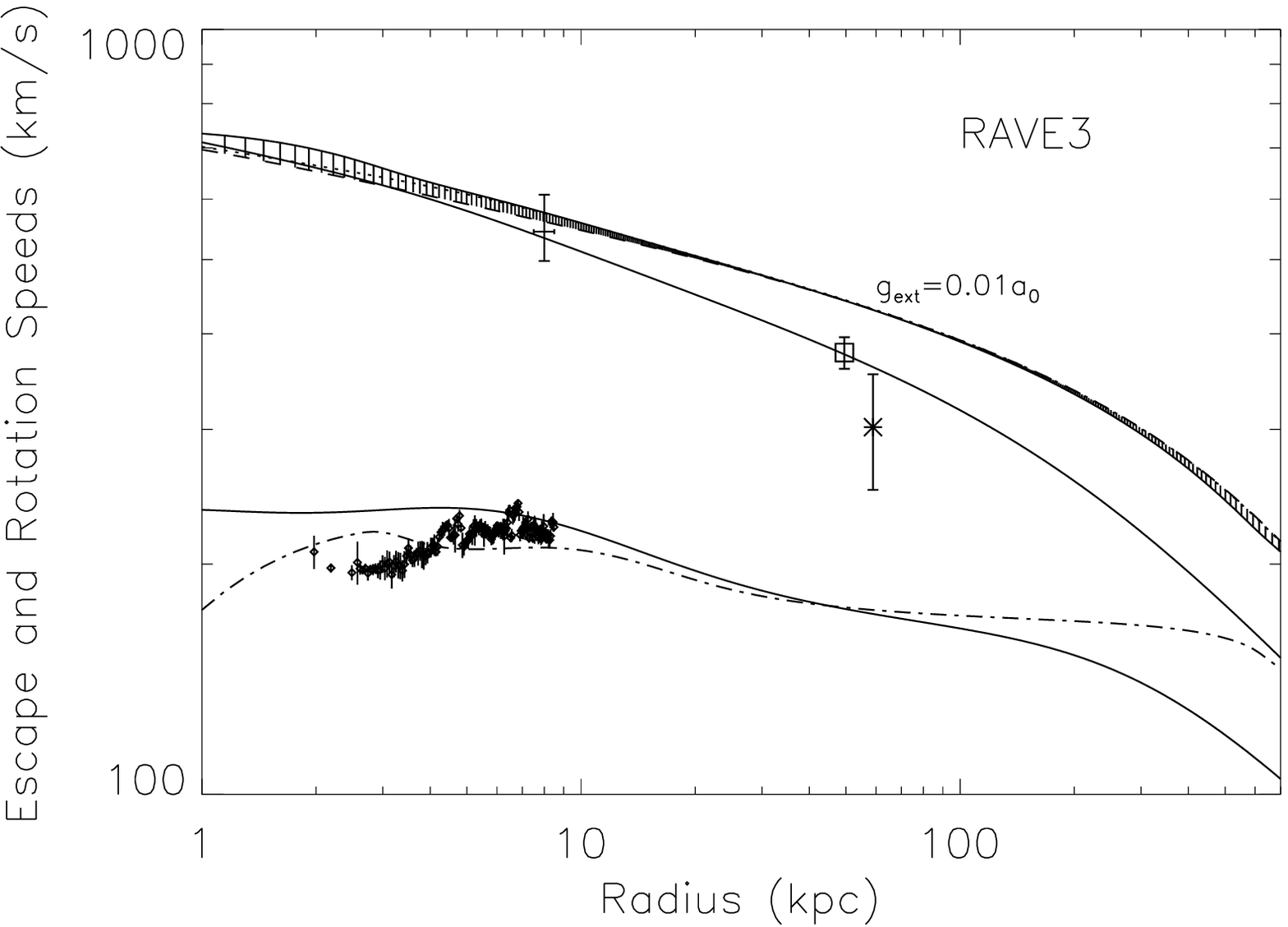}}
\vskip 0.5cm
\caption{RAVE models of the Milky Way. The left and right panels are figures for the NFW and WE halos respectively. The symbols are the same as in Fig~\ref{kzsbsc}.}\label{rave}
\end{figure}

The LMC is the nearest satellite of the Milky Way,  located in galactocentric coordinates at $(x,y,z)=(-0.8 \, {\rm kpc}, -41.5  \, {\rm kpc}, -26.9 \, {\rm kpc})$ (Murai \& Fujimoto 1980; Kallivayalil et al. 2006a,b; Besla et al. 2007). Recent measurements of the proper motion of the LMC with the Hubble Space Telescope have shown that its velocity is as fast as $378\pm18\kms$ (Kallivayalil et al. 2006a). Previous observations suggested the 3D velocity was ranging from $249.3~\kms$ to $ 367.83~\kms$ (Mastropietro et al. 2005; van der Marel et al. 2002; Murai \& Fujimoto 1980; Gardiner et al. 1994; Gardiner \& Noguchi 1996; Heller \& Rohlfs 1994). These previous values of the LMC's proper motion velocity were consistent with it being bound to the Galaxy in CDM models. However, with its newly measured velocity, Besla et al. (2007) already pointed out that it is difficult for the LMC to be a bound satellite, especially if one assumes CDM-based halos rather than an isothermal one (Hernandez et al. 2001), unnatural in the CDM framework (and actually resembling the MOND potential, the external field defining the truncation). The positions and velocities of the Magellanic Cloud system are listed in Table~2, while the escape velocity at the position of the LMC in our different CDM and MOND models are listed in Table~3. It can be directly seen that the CDM models can only weakly bind the LMC, the huge velocity of the LMC being in the critical range of escaping. However, we see that a MOND model with a weak external field of $0.01a_0$ provides a much deeper potential at the galactocentric radius of the LMC and strongly binds the Magellanic Clouds System. The SMC is bound in both the MOND and CDM frameworks, but in CDM it leaves open the question of why the brother galaxy LMC is close to escape. Note,  however, that within one sigma-error, the SMC could also be unbound to the MW in all four CDM models, and that the LMC can still be bound at the one-sigma level in three of our four CDM models.

\begin{table*}
\caption{observed 3D velocity and escape velocity at the position of the LMC (in unit of $\kms$)}
\begin{tabular}{ccccccc}
\hline
Observation & MOND(EF=$0.01a_0$)&MOND(EF=$0.03a_0$) & KZS & Besan\c{c}on & RAVE1 & RAVE3\\
\hline
$378\pm18$ & $(441.8, 442.8, 442.0)$& $(366.0, 367.8, 366.9)$ & $368.9$ &$366.9$ & $348.6$&$375.7$ \\
\hline
\end{tabular}
\end{table*}

The first panel of Fig~\ref{kzsbsc} shows the circular and escape velocity curves (on a logarithmic scale) of the KZS B1 model compared with the MOND (low external field) model. This logarithmic abscissa allows us to have a clearer view of the large-scale behavior of the circular and escape velocity curves. For the CDM model to be consistent with the observed circular velocities and RAVE-inferred escape velocity, the escape velocity at the position of the LMC must be smaller than (or equal to) the newly measured mean-value of the 3D velocity. Besla et al.~(2007) have shown that an escape velocity of $552\kms$ at 50~kpc could still be consistent with the rotation curve of the inner Galaxy, if the virial mass within 200~kpc is increased to $2\times 10^{12}\msun$ and the concentration is decreased to $c=9$.  However, this model is in direct contradiction with the local escape velocity from the Solar neighbourhood measured by Smith et al.~(2007). As for the KZS B1 in which the CDM halo could still bind the LMC, such a case is marginal and the LMC is only weakly bound. In contrast, the MOND potential with an external field of $0.01a_0$ is significantly deeper. Therefore, the LMC is undoubtedly captured by the Galaxy in this MOND potential.

The escape velocity as a function of radius (on a logarithmic scale) for the Bsc model is shown in the second panel of Fig.~\ref{kzsbsc}. Again, the LMC speed is in the critical escape velocity range in the NFW CDM framework. On the two panels of Fig.~\ref{kzsbsc}, the CDM escape velocities are nearly the same, especially at large radii, due to the fact that they have the same CDM halo profile. The baryons affect the rotation curve in the central regions, but have little contribution to the escape velocity in the CDM framework.

Finally, we compare the escape velocity in the two RAVE models with the MOND (low-external field) model in Fig.~\ref{rave}. The problem of binding the LMC to the Galaxy is even more severe in this case; the NFW halo cannot bind the LMC anymore. The RAVE~3 model also overestimates the circular velocity because the central region of the Wilkinson-Evans profile is much too dense. However, even in this RAVE~3 model, the escape velocity at the LMC position is predicted to be similar to the actual velocity of the LMC, which means that the LMC is only bound at the one-sigma level. 

If more precise measurements of the 3D velocity of the LMC in the future imply that the LMC is truly unbound to the Galaxy in the CDM models, it could be a serious problem (i) to explain why the orbital angular momentum of the LMC is pointing 9 degrees away or closer to the mean direction of the orbital poles of the ancient satellites Ursa Minor, Fornax, Draco, and Carina (Metz et al.~2008), and (ii) to explain the formation of the Magellanic stream by tidal stripping or ram pressure stripping. As already advocated by Besla et al. (2007), the LMC would be on a parabolic orbit and would not have suffered any pericentric passage or crossed the disk before. 
In MOND with a low external field the LMC could then still be on a bound orbit, but the problem might be that this orbit would have a very long period (approximately 3~Gyr), and might not cross the disk often enough. However, the effect of the external field of the Galaxy on the self-gravity of the LMC might make it easier for ram pressure stripping to take place. A detailed study of the orbit and evolution of the internal structure of the LMC in MOND is however beyond the scope of this paper, and will be the subject of further studies.

\subsection{Constraining MOND and CDM potentials with Hypervelocity stars}
Measurements of the LMC velocity thus provide a unique quantitative observational tool for studying the Galactic potential at large distances. However, an additional observational measure, which is independent of the LMC velocity, may become accessible in the near future through the observations of hypervelocity stars (HVSs) .
In recent years several HVSs have been observed at large distances in the Galactic halo (up to ~100 kpc; Brown et al. 2007a). These stars have (most likely) been ejected from the Galactic centre with high velocities, probably following binary disruption by the massive black hole in the Galactic centre (Hills 1988; Yu \& Tremaine 2003; Perets 2007; Perets et al. 2007). Some of these HVSs have velocities extending much beyond the escape velocity from the Galaxy, where others are still bound to the Galaxy (Brown et al. 2007b). Recently Gnedin et al. (2005) and Yu \& Madau (2007) suggested to use the kinematics of HVSs in order to probe the galactic potential using the position and velocity vectors of HVSs at large Galactocentric distances. Under the assumption that HVSs were ejected from the galactic centre they suggest to measure the departure from purely radial orbits of these HVSs, due to the (possible) triaxiality of the galactic potential. The slight perturbations from radial orbits are then used in order to estimate the level of triaxiality of the Galaxy. Both Gnedin et al. and Yu \& Madau have focused on measuring the triaxiality of the galactic potential in the context of CDM models. However, given the large deviation of MOND potentials from spherical symmetry (relative to CDM potentials), these methods could be even more useful for exluding or confirming MOND models for the Galaxy. In addition, studies of propagation times and especially the return times of bound hypervelocity stars (Perets et al. 2008 in preparation) could be highly valuable in constraining MOND potentials such as discussed here up to very large distances ( $>$100 kpc). More measurements of HVSs velocities, especially at large galactocentric distances are expected to be obtained in the few coming years, and could much better constrain our knowledge on the galactic potential.

\subsection{Escape velocity with a component of hot baryons}

For completeness we explore the sensitivity of our predictions to the possibility of a small amount of uncooled, moderately hot baryons existing at large radii in the Galaxy. For this purpose, we added to the Besan\c{c}on (cool) baryons model a Plummer sphere of $4\times 10^{10} M_\odot$ with a Plummer scale-length of 100~kpc. In CDM models, this does not affect the prediction since the potential is completely dominated by the CDM halo. In MOND, however, the prediction becomes quite different. The rotation curve remains almost unchanged in both the $0.01a_0$ and $0.03a_0$ external field cases, but the escape velocity is boosted by about 30~km/s at the Sun's position. In that case, the $0.01a_0$ external field model is only marginally compatible with the RAVE-measured local escape velocity, while the $0.03a_0$ model becomes more comparable to the original $0.01a_0$ model, and stronly binds the LMC while still being compatible with the local escape velocity (see Fig.~7). This again demonstrates that MOND models are not as unique as commonly assumed.

\begin{figure}{}
\resizebox{8cm}{!}{\includegraphics{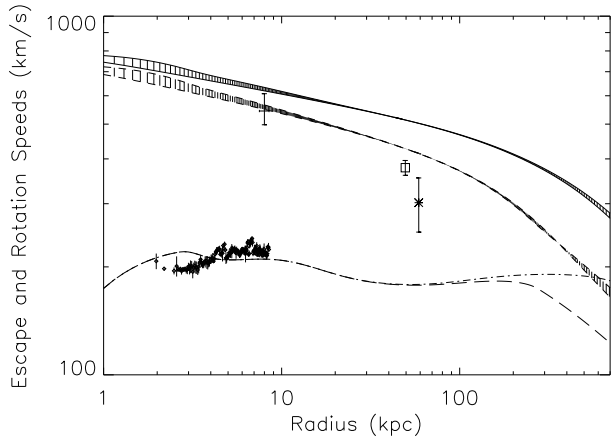}}
\vskip 0.5cm
\caption{Circular and escape velocities in Besan\c{c}on MW baryon model + hot, diffused gas around MW. A Plummer model for the hot gas component is used, with mass of $4\times 10^{10}\Msun$. The two shaded areas are the escape velocities within external field of $0.01a_0$(upper) and $0.03a_0$(lower), and the two bottom curves show the circular velocities. The dot-dashed line is the MOND rotation curve for the MW embedded in an external field of $0.01a_0$, and the long dashed line is for the $0.03a_0$ external field. The symbols of observations are the same as in Fig~\ref{kzsbsc}.}\label{rave2}
\end{figure}

\section{Conclusion \& Discussion}

In this paper, we modelled the Milky Way with several different CDM-based models, and compared them with the MOND framework. We first showed that the isopotentials are more spherical in CDM, at small radii because the baryonic matter does not affect the potential as much as in MOND, and at large radii because the external field flattens the potential in MOND. In CDM, the halo dominates the shape of the potential at large radii, and consequently the escape velocity at different positions on a given spherical shell show only negligible differences. Note that all of our studied CDM halos were spherical, while recent studies on CDM found that the density profiles of massive halos should rather be triaxial (Jing \& Suto 2002; Bailin \& Steinmetz 2005; Allgood et al. 2006; Andrea et al. 2007). A mean axis ratio $\bar q={a_2+a_3\over2a_1}$ is defined, where $a_1$, $a_2$ and $a_3$ are the major, intermediate and minor axis. In Andrea et al. (2007), for a Milky Way-like dark halo with a virial mass of $10^{12}\Msun$, one has $\bar q\sim 0.76^{+0.1}_{-0.15}$. Such triaxial dark halos would change the potential contours inside the virial radius: the potentials would be flattened as well. Nevertheless, when the radius is far larger than the virial radius (e.g. 10 times the virial radius), the potential contours would become spherical again. 

We also showed that the effective, phantom dark matter densities predicted by MOND can be negative at large radii (where the internal and external acceleration are of the same order of magnitude), in a cone perpendicular to the external field (see also Milgrom 1986b). A negative convergence parameter could thus in principle be observed in the gravitational lensing generated by Milky Way-like galaxies, although it would be very sensitive to the detailed non-constant gravitational field of the environment.

In addition,  we  found the circular and escape velocity as a function of position in the different models. We showed that the rotation curves in all models are compatible with the inner rotation curve (Clemens 1985) as well as with the SDSS-measured rotation curve at large radii (Xue et al. 2008). This is true for various $\mu$-functions in MOND (``standard" and ``simple"), but the form adopted here fits best the rotation curve (the ``simple" form of Famaey \& Binney 2005). On the other hand, the prediction of escape velocity as a function of position will be very useful for comparison with inferred  velocities from the data of planned or already-underway surveys such as RAVE, SDSS, SEGUE, SIM or GAIA or the hypervelocity stars survey. We stress that MOND models are far from unique, and that the freedom allowed for the modulus of the external field makes the MOND models actually more flexible than CDM-based ones with respect to the escape velocities that could be derived from the kinematic surveys. High escape velocities would rule out the four CDM models studied here, while any escape velocity between the two curves plotted on Fig.~4 would only represent a constraint on the external field in which the Milky Way is embedded in MOND. We also note that the MOND model with an external field of $0.03a_0$ mimicks the CDM ones in terms of both circular and escape velocity curves. Let us finally point out that we did not take into account any effect of the cosmic acceleration in either the CDM or MOND context, which could have some effect on both predicted escape speed curves.

We then note that the newly-measured value of the 3D velocity of the LMC (Kallivayalil et al. 2006a,b; Besla et al. 2007) could already be an indication that escape velocities in the outer Galaxy could actually be higher than that predicted by the CDM models (as well as by an external field of $0.03a_0$ in MOND), under the assumption that the LMC is bound to the Milky Way. More precise measurements of the 3D velocity of the LMC will be needed to disentangle this issue, but with its present value we found that the LMC is only bound at the one-sigma level in three of our four CDM models (and unbound at the one-sigma level in the fourth one).
In this context, we note that the reanalysis of the LMC proper motion by Piatek et al.~(2007) yields a smaller value of $358\kms$ (radial and tangential velocities then being $93.2\kms$ and $346\kms$), which is consistent with the lower error bound of Kallivayalil et al. (2006a). In the MOND context, the measured velocity of the LMC can be used as a constraint on the modulus of the external field in order to keep it bound: we found that an external field of less than $0.03a_0$ is needed, an upper limit strikingly similar to that needed to explain the local escape speed from the solar neighbourhood (Famaey et al. 2007c; Wu et al. 2007).
We also suggest that future observations of hypervelocity stars could add additional independent constraints that could even better constrain the models studied here.

It is finally important to note that because of their cuspiness, it will always be difficult for the four CDM models explored in this paper to reproduce the non-axisymmetric motions of gas in the longitude-velocity diagrams of the baryon-dominated inner Milky Way, as extensively discussed in Famaey \& Binney~(2005). It remains to be seen what the prediction of a MOND model is for these non-axisymmetric gas motions. Still, since we showed that the MOND potential (and phantom dark matter) tracks the baryonic distribution quite well at small radii, and since the best fit to these non-axisymmetric motions of the gas produced by the bar was obtained with a purely baryonic model (Bissantz et al. 2003), it is likely that they could be explained within the MOND context.

We thus conclude that the MOND model studied in this paper, with an external field of modulus $0.01 a_0$ directed in the Sun-Galactic centre direction, fits all the present observations of the Milky Way extremely well. If a component of unseen, moderately hot, baryons is added, the preferred value of the external field however becomes higher ($0.03a_0$).

\section{acknowledgments}
We thank Luca Ciotti, Pasquale Londrillo, Carlo Nipoti for generously sharing their code, and Garry Angus, Francoise Combes, Olivier Tiret and Spyridon Talaganis for comments. XW acknowledges the support of SUPA studentship. HSZ acknowledges partial support from UK PPARC Advanced Fellowship and National Natural Science Foundation of China (NSFC under grant No. 10428308). BF acknowledges the support of the FNRS.

\bsp

\label{lastpage}

\end{document}